\documentclass{iopjournal}
\usepackage{lmodern}
\usepackage{float}
\usepackage{cite}
\usepackage{siunitx}
\usepackage{booktabs}
\DeclareSIUnit{\ppm}{ppm}
\DeclareSIUnit{\torr}{Torr}

\begin{document}

\articletype{Paper} 

\title{Nb$_3$Sn Films Exhibiting Continuous Supercurrent Across a Diffusion Bonded Seam}

\author{Andre Juliao$^{1*}$\orcid{0000-0003-4350-9977}, Wenura Withanage$^2$\orcid{0000-0003-1350-9423}, Nikolya Cadavid$^{1,3}$, Anatolii Polyanskii$^1$ and Lance D Cooley$^1$\orcid{0000-0003-3488-2980}}

\affil{$^1$Applied Superconductivity Center, National High Magnetic Field Laboratory, Tallahassee, FL 32310, USA}

\affil{$^2$Department of Physics, Florida State University, Tallahassee, FL 32306, USA}

\affil{$^3$Florida A\&M University--Florida State University College of Engineering, Tallahassee, FL 32312, USA}

\affil{$^*$Author to whom any correspondence should be addressed.}

\email{arj14c@fsu.edu}


\begin{abstract}

Multiple pairs of bronze pieces were joined along a common seam and then exposed to Nb vapor via sputter deposition during heating at $\sim$\qty{715}{\degreeCelsius} to form a diffusion bond between the pieces. Polishing and alignment of the pieces created smooth surfaces normal to the Nb flux with seams perpendicular to the surface (i.e. parallel to the Nb flux). Conversion of Nb to Nb$_3$Sn took place simultaneously with diffusion bonding, resulting in Nb$_3$Sn thin films that coated bronze surfaces and spanned seams with uniform thickness. Characterization of superconducting properties via magneto-optical imaging suggests that supercurrent flows freely across the seam in several examples when cooled to \qty{9}{\kelvin} and shielding or trapping low magnetic field.
Modification of the process to coat the pieces with Nb prior to diffusion bonding and Nb$_3$Sn formation resulted in varying degrees of seam coverage by the resultant Nb$_3$Sn films. The pre-coating method did not produce any example with quality comparable to the examples obtained by the hot bronze approach.
This work could enable new approaches to joining Nb$_3$Sn materials in magnet conductor and RF cavity applications.
\end{abstract}

\section{Introduction}


 Joints are an essential component of superconducting technologies. High-field nuclear magnetic resonance (NMR) magnets depend on various configurations to form persistent joints between magnet conductors, including spot welding, diffusion bonding, pressure contacts, and use of intermediate superconductors such as solder \cite{Schneider1997, Swenson1999, Brittles2015, Maeda2019, Speller2022}. For intermetallic superconductors such as Nb$_3$Sn, spot welding, pressure contacts, and diffusion bonding are not practical, and joints are presently made using a secondary superconducting material and locating the joined region away from high magnetic field \cite{Brittles2015}.

Superconducting radio-frequency (SRF) cavities are also made by joining components. Most cavities are constructed by welding Nb pieces as reviewed in \cite{Singer2016}, which creates a pristine connection that accommodates RF currents operating at a significant fraction of the depairing current. The only present alternative to welding is to use seamless technologies \cite{seamless} or coatings, such as the Nb coatings applied to Cu for the Large Electron-Positron Collider \cite{LEP1, LEP2}. Research and industry needs have progressed to emphasize Nb$_3$Sn coatings as the present state of the art \cite{Posen2017, Dhuley2020, Ciovati2020}. The primary method of cavity scale Nb$_3$Sn coatings has been mainly limited to the vapor diffusion technique, where Sn vapor is reacted with a bulk Nb cavity \cite{Posen2017}. This route achieves quality factor high enough to consider operation of large scientific facilities with increased economy, as well as enable commercial operations with cryocoolers \cite{Dhuley2020, Ciovati2020, Stilin2020}. As with many emerging technologies, connections between macroscopic processing and details at the micro-scale that limit performance are still emerging, see for example \cite{Posen2017} and \cite{Lechner2024} which discuss different states of development separated by nearly a decade.

Many groups are directing effort to solve materials challenges connected with Nb$_3$Sn SRF cavities that do not need reaction temperatures near or above the melting point of copper, \qty{1085}{\degreeCelsius}, where Sn-vapor methods are carried out. Our group is one of several that are considering bronze as both a base material onto which Nb is deposited \cite{Withanage2021,Juliao2026,Zhu2022,Rey} and a material covering Nb that is removed after the reaction is complete \cite{Juliao2026,MLu2022, MLu2025}. Direct deposition from a Nb$_3$Sn source \cite{Ilyina2019} and formation of Nb$_3$Sn via chemical vapor deposition \cite{Gaitan} are also promising routes.

Effort in non-superconducting RF cavities for linear accelerators and astronomic detectors relies upon forming the cavity body as separate pieces, which are joined after intensive material preparation and inspection steps \cite{Alesini2019, Golm2022, Braine2023, Posen2023}. These ``clamshell'' approaches align seams with RF current flow in the primary cavity mode to reduce losses and flux admission at the seam. Clamshell approaches for SRF cavities have attractive features such as accommodation of line-of-sight deposition methods and direct inspection and feedback from the cavity surface during processing.


In this article, we extend previous work discussed in \cite{Withanage2021} that explored formation of Nb$_3$Sn with the assistance of bronze (Cu-15wt.\%Sn). We investigate whether a continuous Nb$_3$Sn layer can be created across the seam between bronze pieces using synthesis routes in the prior work. We compare a novel Nb$_3$Sn growth technique, the ``hot bronze'' method, with traditional post-reaction approaches to coat two joined bronze pieces. The hot bronze method applies Nb via magnetron sputtering onto a Cu-15wt.\%Sn bronze substrate held at $\sim$\qty{715}{\degreeCelsius} to facilitate instantaneous formation of Nb$_3$Sn. This method was applied decades ago to form very small grains of Nb$_3$Sn for flux-pinning studies \cite{Cooley2002}. We present extensive analyses of the microstructure and microchemistry resulting from the different processing approaches.

In addition to checking the film morphology, we examined the superconducting properties using magneto-optical imaging (MOI), which is exquisitely sensitive to blockage of supercurrent by defects and boundaries. MOI characterizations gave a clear indication of whether the seam location might have adverse impact on applications such as detectors. 

\section{Experimental detail}

\begin{figure}[bt]
        \centering
        \includegraphics[width=.6\textwidth]{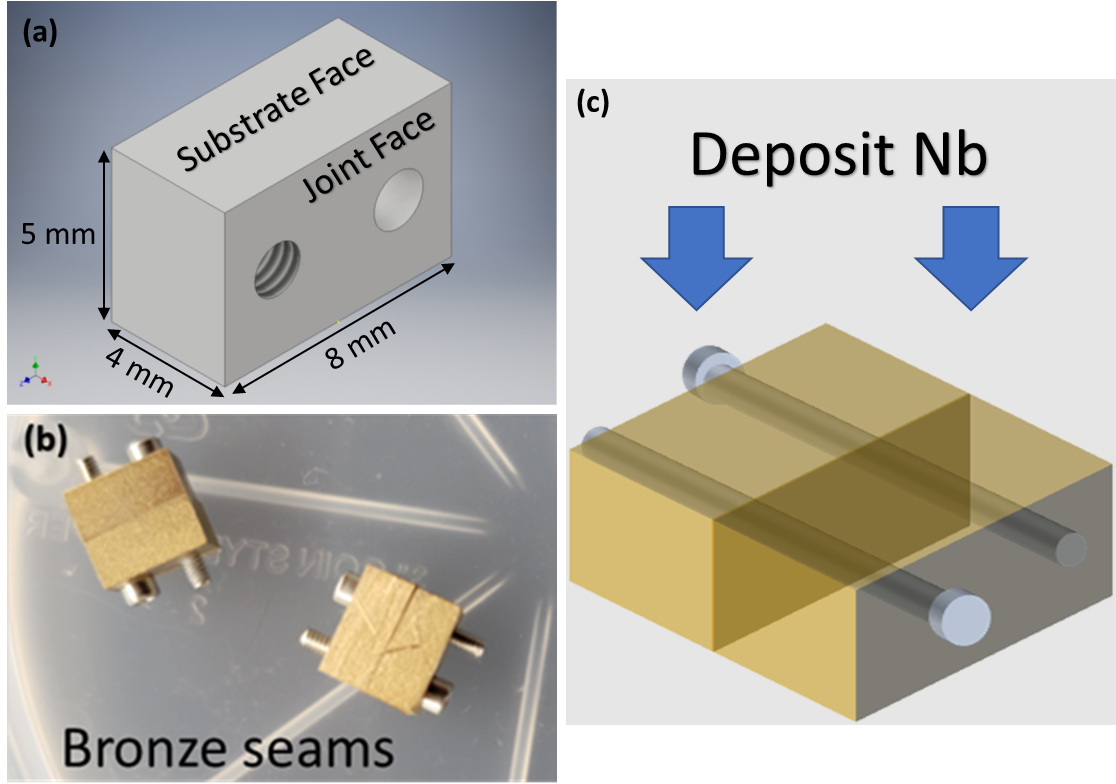}
        \caption[]{Schematic (a) shows a diagram of one bronze piece used in this experiment. The block contains one clear and one tapped hole for bolting. The faces that were polished are also identified. In photo (b), two-block assemblies are shown bolted together. Schematic (c) illustrates how Nb deposition was applied to two bolted pieces forming a joint. Not shown are alternative methods where Nb was deposited on individual blocks before bolting them together. }
        \label{fig:BronzeBlockSchem}
    \end{figure}

In this work, we used
Cu-15wt.\%Sn (Cu-9at.\%Sn bronze, a high Sn $\alpha$-bronze phase, in reference to the Cu-Sn phase diagram)
as substrate materials. As explained in more detail in our previous work \cite{Withanage2021}, this bronze substrate was prepared at the Ames Laboratory.
Blocks of each material with approximate dimensions \qtyproduct{8 x 4 x 5}{\milli\meter} were prepared as shown in Fig.~\ref{fig:BronzeBlockSchem}(a). Pairs of blocks were then drilled with alignment holes to facilitate clamping with stainless steel screws.

Two faces of each block were polished to a mirror finish. One of these faces was pressed against an equivalent polished face of the second block in the pair; this is called the joint face. The other face became the substrate for Nb deposition and Nb$_3$Sn formation; this is called the substrate face. These faces are illustrated in Fig.~\ref{fig:BronzeBlockSchem}(a). Fig.~\ref{fig:BronzeBlockSchem}(b) shows assembled, unpolished block pairs with clamping screws. The joint faces were clamped together and the clamped assembly was subsequently polished to produce a mirror finish for the substrate face as illustrated in Fig.~\ref{fig:BronzeBlockSchem}(c). The average roughness ($R_a$) of the mirror-finished bronze surfaces was determined using an atomic force microscope (AFM). Scans indicated a typical roughness of \qty{2}{\nano\meter} on a \qty{25}{\micro\meter\squared} scale. Height variation across the joint was \qty{<10}{\nano\meter} in the best joints, and as high as \qty{25}{\micro\meter} for joints that were unbolted, Nb-coated, then rebolted.

Cleaning polished blocks or block assemblies used a sequence of steps with deionized water, acetone, and isopropyl alcohol in an ultrasonic bath. Drying took place with compressed air before the blocks were loaded into the deposition chamber. The load lock provides heating to \qty{110}{\degreeCelsius} in high vacuum, which was used to remove volatile surface residues over a 10-minute period before introduction into the main chamber. All heat treatments and sputter depositions were performed in an ultra-high vacuum (UHV) sputtering chamber with a base pressure of \qty{e-9}{\torr}. Nb was deposited using a 2-inch magnetron gun at \qty{250}{\watt} with a \qty{8}{\milli\torr} Ar working gas, with a distance of approximately \qty{40}{\centi\meter} between the gun and substrate. Pre-sputtering of the guns took place behind a closed shutter for 3 minutes to remove surface contaminants from the target prior to deposition. Substrate heating was provided by a resistive heater embedded in the substrate support, with thermocouples used for temperature monitoring and control. Calibration runs were performed to correlate thermocouple readings with the actual bronze substrate temperature. 

Four different processing recipes are
outlined below. The intention of the recipes was to isolate different compositional and structural changes associated with diffusion bonding of the joint and growth of the Nb$_3$Sn coating across the seam. Deposition temperature of \qty{200}{\degreeCelsius} was used to obtain smooth Nb films without any diffusion reaction. Temperature of $\sim$\qty{715}{\degreeCelsius} drove a diffusion reaction between Nb and Sn held in the bronze to form Nb$_3$Sn. Each recipe aimed to yield a Nb$_3$Sn thickness of $\sim$\qty{1}{\micro\meter}.

\begin{itemize}
    \item \textbf{Recipe CS} for cold Nb deposition on separate blocks: After polishing the substrate face of a clamped block pair, the bronze pieces were unbolted and coated separately with Nb at \qty{200}{\degreeCelsius} through a mask to conceal the joint face. Blocks were then reassembled and re-bolted, introduced through the preparation chamber, and reacted in the main deposition chamber at $\sim$\qty{715}{\degreeCelsius} for 4 hours in UHV. This method was not successful.

    \item \textbf{Recipe CJ} for cold Nb deposition on joined blocks: After polishing the substrate face of a clamped block pair, the bolted assembly was introduced into the preparation chamber and Nb was deposited at \qty{200}{\degreeCelsius} and immediately followed by a 4-hour reaction at $\sim$\qty{715}{\degreeCelsius} in the deposition chamber. This method was not successful.

    \item \textbf{Recipe CB} for cold Nb deposition on diffusion-bonded blocks: After polishing the substrate face of a clamped block pair, the bolted assembly was introduced into the preparation chamber. The bolted bronze blocks were pre-heated at \qty{715}{\degreeCelsius} for 2 hours to facilitate diffusion bonding. The assembly was then cooled to \qty{200}{\degreeCelsius} and Nb was deposited, then the temperature was again raised to $\sim$\qty{715}{\degreeCelsius} for 4 hours while the assembly was still in the deposition chamber. This method gave mixed results.

    \item \textbf{Recipe HB} for hot Nb deposition on diffusion-bonded blocks (related to the ``hot-bronze'' method described earlier): After polishing the substrate face of a clamped block pair, the bolted blocks were pre-heated at $\sim$\qty{715}{\degreeCelsius} for 2 hours. Nb was then deposited with the bronze still at \qty{715}{\degreeCelsius}, resulting in immediate formation of Nb$_3$Sn. Deposition was followed by continuing to hold temperature at \qty{715}{\degreeCelsius} for 1 hour before cooling in the chamber. This method was successful.

\end{itemize}

The superconducting critical temperature of the samples was extracted from measurements of magnetic moment ($m$) vs. temperature ($T$) curves. The critical temperature $T_c$ onset was identified as the highest temperature at which the normalized magnetic moment deviates from zero, indicating the onset of the diamagnetic Meissner response. Data was acquired using a SQUID magnetometer (Quantum Design MPMS-5) by mounting samples with the film plane parallel to the scan axis and applied magnetic field. Measurements used a zero-field-cooled (ZFC) method where, after removing the magnet hysteresis using a standard oscillatory approach, the sample was first cooled to $\sim$\qty{4}{\kelvin} in zero-field, then a \qty{1}{\milli\tesla} field was applied and held constant while recording data at small increasing temperature increments up to \qty{20}{\kelvin}. 

Surface morphologies were analyzed using an FEI Helios G4 UC field emission scanning electron microscope (SEM) in secondary electron (SE) imaging mode. The elemental compositions were assessed via energy dispersive x-ray spectroscopy (EDS) using an Oxford Instruments X-MaxN SDD X-ray detector (standard-less analysis). Care was taken to limit influences from regions adjacent to the electron beam when acquiring spectroscopy data. Surface roughness and topographical information were measured using a Veeco Icon atomic force microscope (AFM). Magneto-optic imaging of the samples was carried out by using a custom setup described in prior literature \cite{Polyanskii2021}.


\section{Results and Discussion}

A summary of the recipes and their results is given in Table~\ref{tab:recipes}. Recipe HB successfully formed a continuous Nb$_3$Sn film bridging the two Cu-Sn bronze substrates.


\begin{table}[htb]
    \centering
    \caption{Summary of the fabrication recipes.}
    \begin{tabular}{lp{\dimexpr\textwidth-2cm}}
        \toprule
        Recipe & Comments \\
        \midrule
        CS & Misalignment at the joint, peeled Nb$_3$Sn, Gaps between joint faces \\
        CJ & Gaps between joint faces, separation of Nb$_3$Sn from the substrate \\
        CB & Evidence for diffusion bond of the joint and continuous Nb across the joint, but discontinuous connection between Nb$_3$Sn and the substrate\\
        HB & Evidence for diffusion bond and continuous Nb$_3$Sn across the joint well-bonded to the substrate.\\
        \bottomrule
    \end{tabular}
    \label{tab:recipes}
\end{table}

\subsection{Challenges with cold Nb deposition (CS, CJ, CB)}


The inclusion of recipes with a \qty{200}{\degreeCelsius} Nb deposition aimed at avoiding a potential process where a magnetron sputtering source would operate inside a hot cavity, effectively acting as an oven. Formation of Nb$_3$Sn occurred during a post-deposition reaction at \qty{715}{\degreeCelsius}. This choice exposed vulnerabilities associated with the large mismatch of the coefficient of thermal expansion (CTE) between different materials, where Nb and Nb$_3$Sn have CTE values of about \qty{9}{ppm}, whereas coefficients for Cu and bronze are about \qty{16}{ppm} \cite{Easton1980}. As the substrate expands during the heating process, it is possible to break the Nb film if adhesion is poor and if the deposition did not impose sufficient residual compressive stress to counteract the tensile stress applied by the substrate during heating. We chose deposition parameters that should not produce large residual stresses in the Nb film, so disruption of the Nb film during warm-up to the reaction temperature is plausible.

\begin{figure}
    \centering
    \includegraphics[width=\textwidth]{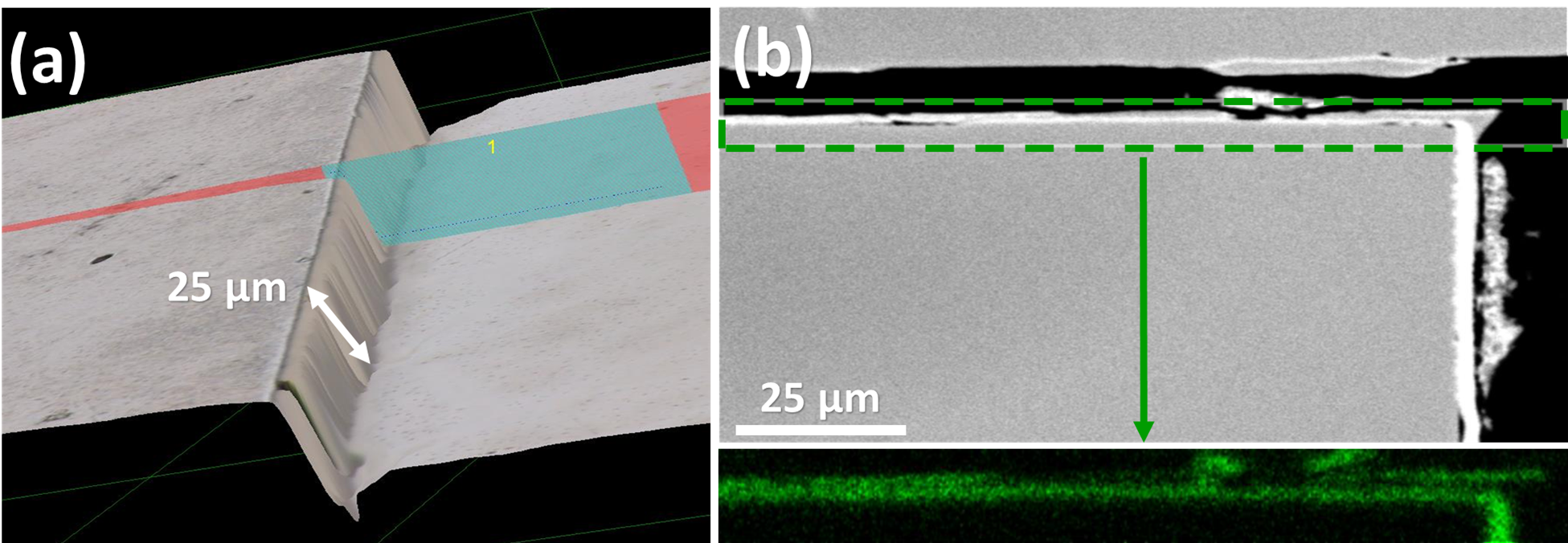}
    \caption[]{Challenges associated with joining separately coated blocks and using a cold-deposition are demonstrated in this figure. In (a), the misalignment between two blocks coated with Nb using the CS recipe is shown. 
    In (b), a cross-section of the same block reveals Nb coverage on both the substrate face and the joint face. Nb at the joint interface (confirmed by EDS mapping shown in green below (b)) was observed in all cold-deposition recipes (CS, CJ, and CB). }
    \label{fig:ChallengesTogether}
\end{figure}

The CS recipe was explored to simulate a process in which the halves of a clam-shell cavity are coated and inspected before the pieces are closed and a high-temperature reaction to form Nb$_3$Sn. Fig.~\ref{fig:ChallengesTogether}(a) shows how the efforts here produced a \qty{25}{\micro\meter} misalignment at the joint, which greatly exceeds the \qty{1}{\micro\meter} film thickness. Clam-shell cavities explored in the literature \cite{Braggio} use alignment pins and other techniques to assure that the joined seam is smooth. None of these precautions were taken in the present experiment. This experiment also revealed evidence for peeling of the coating at the joint and infiltration of Nb onto the joint faces of blocks as highlighted in Fig.~\ref{fig:ChallengesTogether}(b). Infiltration of Nb obstructed the inter-diffusion of Cu and Sn across the joint face, reducing or preventing the formation of a good diffusion bond.

\begin{figure}
    \centering
    \includegraphics[width=\textwidth]{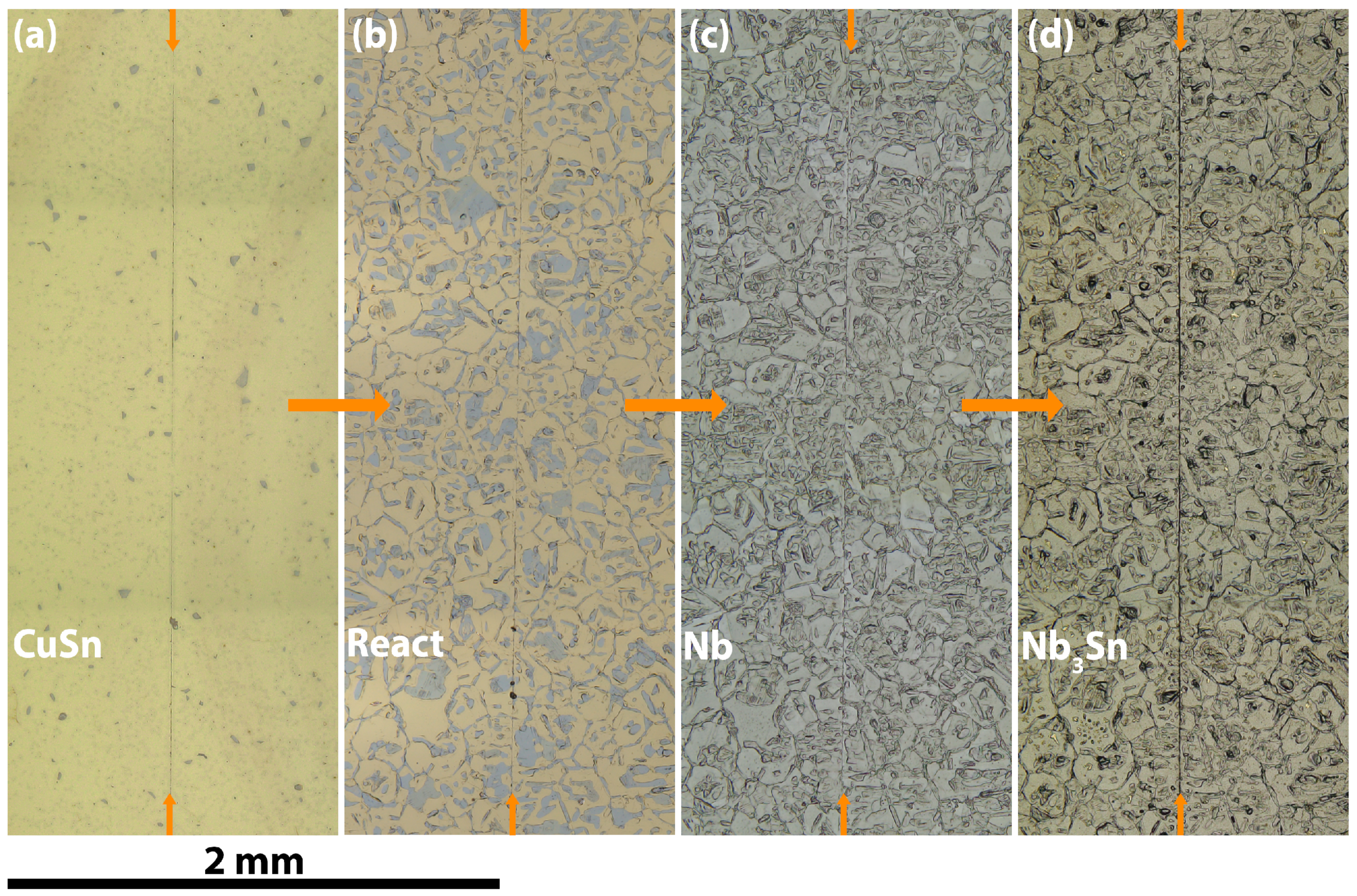}
    \caption[]{These four panels show the same location of the same block at sequential stages of the CB recipe. Orange arrows top and bottom in each panel guide to the seam. In (a) bronze regions appear golden in color and regions with excess Sn appear as small gray regions. The joint seam is evident with the \qty{1}{\micro\meter} gaps at different locations along the joint. Photo (b) is taken after the \qty{715}{\degreeCelsius} dwell for 3 hour step, aiming to produce a diffusion bond between the bronze blocks. The bronze (low-Sn $\alpha$ Cu-Sn phase) is copper colored. Since the Cu-Sn phase diagram allows for a reduction of Sn content in the $\alpha$ phase at the reaction temperature, the light gray blocks represent a quenched-in Sn-rich phase that we identify as $\beta$. In some regions, grains have recrystallized across the joint, suggesting good bonding, whereas in other regions the seam is still evident. In (c), Nb is deposited at \qty{200}{\degreeCelsius} covering the surface scene in (b). The Nb film takes on the underlying bronze grain structure, and artifacts of the $\beta$ phase are evident. The seam is less visible, suggesting nearly complete coverage of the joint. Image (d) shows the surface after reaction at \qty{715}{\degreeCelsius} for 4 hours to convert Nb into Nb$_3$Sn. Again, the underlying bronze grains are evident. The seam profile is more visible than in previous steps. In (c) and (d), the magnification is too low to reveal the nanoscale Nb and Nb$_3$Sn grains.}
    \label{fig:Recipe3Seam}
\end{figure}

The CJ and CB recipes aimed to simulate a process in which bronze pieces are first joined and then Nb is applied to the substrate face at \qty{200}{\degreeCelsius} with a subsequent post-reaction converting the Nb to Nb$_3$Sn. For a cavity, this could be accomplished using independent heating steps to ensure the magnetron is not operated in an oven-like environment. Figure~\ref{fig:Recipe3Seam} shows a sequence of panels looking at the substrate face across the joint at successive stages in the CB process. In (a), the joint is visible between the bolted blocks due to $\sim$\qty{1}{\micro\meter} gaps between the blocks. 
Image (b) suggests bonding has occurred along much of the joint, as evidenced by the presence of grains that span the seam. Regions of excess Sn are observed, as described in the figure caption. After Nb deposition, image (c), the joint was difficult to discern and the film appeared to be mostly continuous. However, after reaction to Nb$_3$Sn the seam once again became more clearly visible as shown in (d).

\begin{figure}
    \centering
    \includegraphics[width=\textwidth]{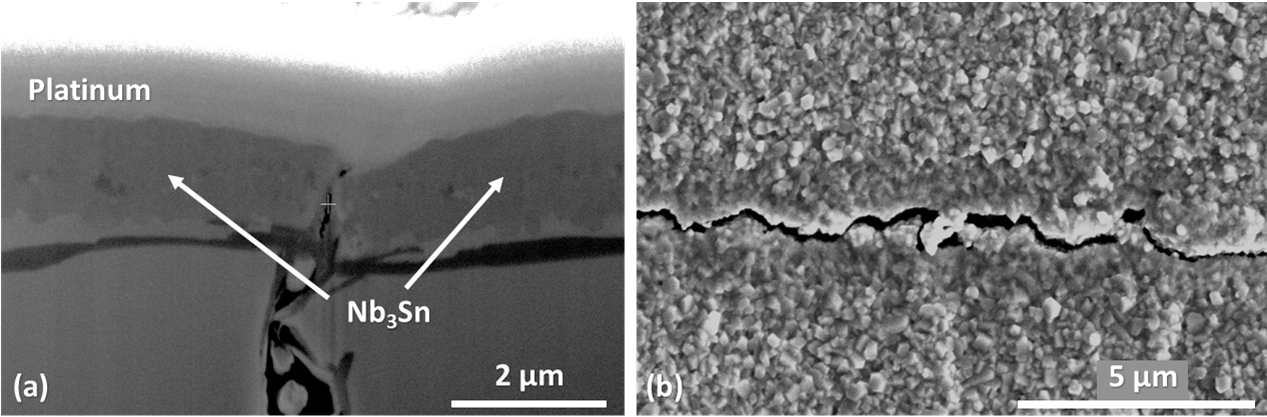}
    \caption[]{Micrograph (a) shows a transverse cross-section of the joint formed by the CB recipe. The joint gap runs vertical in the center of this image, while the Nb$_3$Sn film runs horizontal across the image. The sample has been covered with a protective Ag layer prior to polishing. Micrograph (b) shows a plan view of an adjacent location, looking down onto the Nb$_3$Sn layer. The joint runs horizontally across this image.}
    \label{fig:CB-Cross-section}
\end{figure}

A cross-section image of the final joint using the CB recipe is shown in Fig.~\ref{fig:CB-Cross-section}(a). The region selected shows the \qty{1}{\micro\meter} gap between the bronze blocks noted in Fig.~\ref{fig:Recipe3Seam}(a), where evidently Nb infiltrated the gap and formed debris in the joint interface. The Nb$_3$Sn film appears to have been cracked after its initial formation above the joint, as indicated by the vertical crack in micrograph Fig.~\ref{fig:CB-Cross-section}(a). This interpretation is supported by the plan view micrograph Fig.~\ref{fig:CB-Cross-section}(b), where an uneven crack runs across the Nb$_3$Sn layer above the joint gap. Also evident is delamination of the Nb$_3$Sn layer from the bronze. Curiously, the delamination does not seem to run along the Nb$_3$Sn-bronze interface, and rather about \qty{100}{\nano\meter} below into the bronze. This suggests excellent adhesion of the Nb$_3$Sn film to the bronze. Together, the micrographs suggest that more successful coatings might be obtained by closing the joint gap and ameliorating CTE mismatch, e.g. by supporting the bronze with a material with lower CTE.

\subsection{Healed seams with Nb deposition onto hot bronze (HB)}

\begin{figure}
    \centering
    \includegraphics[width=.8\textwidth]{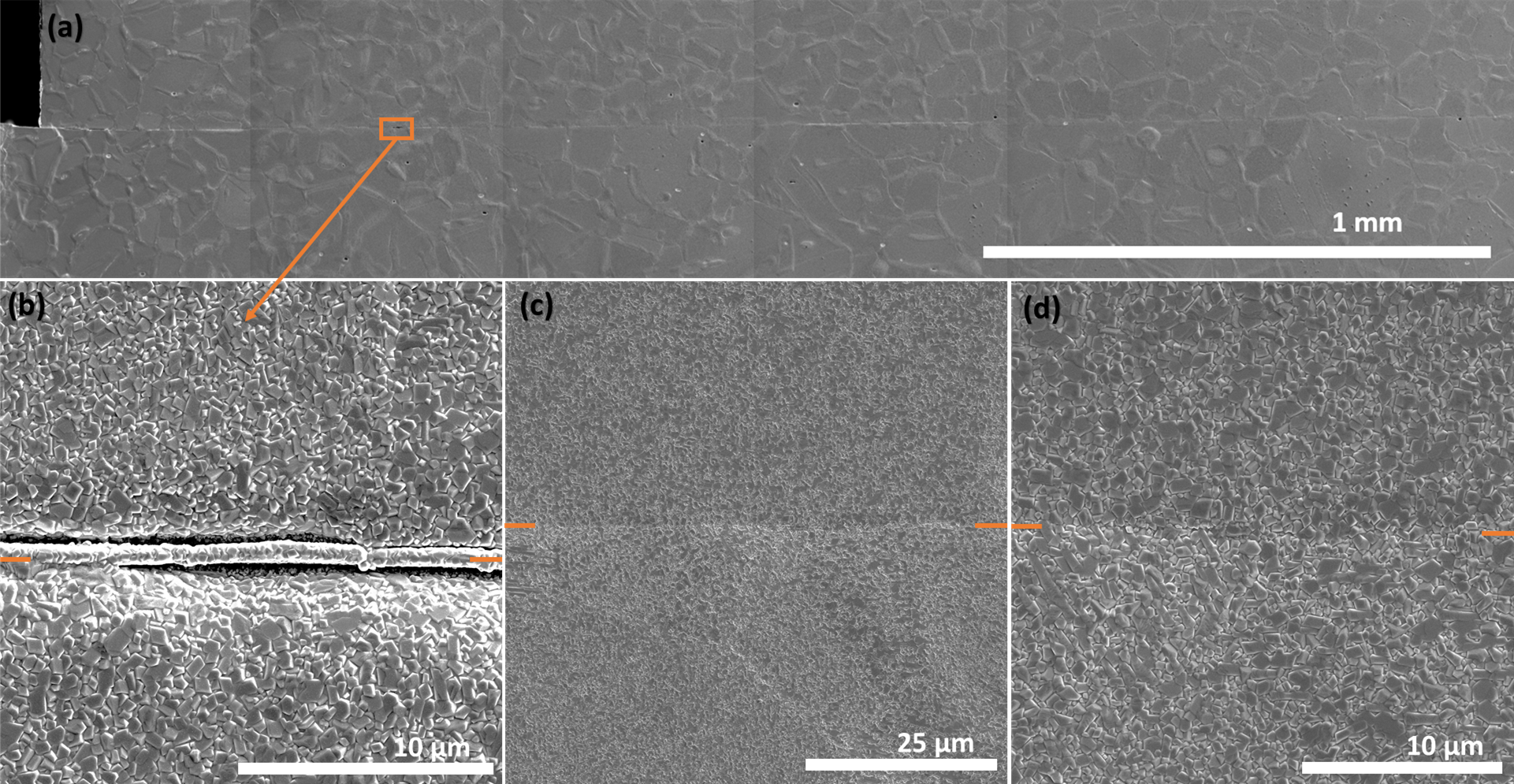}
    \caption[]{The stitched image in (a) shows a plan view looking down on the Nb$_3$Sn coating deposited on the mutual substrate face of joined bronze blocks. Grains in the underlying bronze are evident due to height differences at the grain boundaries. The seam between the blocks is also evident by the same height contrast mechanism, which is similar in contrast to that caused by the underlying bronze grains. In (b), (c), and (d), magnified views of different areas of the joint are shown. A rare defective area highlighted by the orange box in (a) is magnified in (b), showing a local gap in the joint with Nb$_3$Sn grains partly adhered to each side. Panels (c) and (d) show completely healed regions in the seam where Nb$_3$Sn grains have grown continuously across the seam. The grains mostly conceal the seam in panel (d). The vast majority of the seam was completely healed. Orange arrows in each panel mark the location of the joint.}
    \label{fig:WenuraZoom1}
\end{figure}

The HB recipe, which could require the operation of a magnetron source inside of an oven-like heated cavity, created a continuous superconducting film spanning the seam. In plan view looking down onto the substrate face at low magnification, height contrast is evident from both the underlying recrystallized bronze grains as well as the joint, as shown in Fig.~\ref{fig:WenuraZoom1}(a). While the seam between the joint faces is clearly evident, its contrast is similar to that of the underlying bronze grains in most areas, suggesting that height differential at the seam is comparable to that in the material bulk on either side. At higher magnification in plan view, Nb$_3$Sn film grains are observed growing continuously along the common substrate face across the joint, as shown in Fig.~\ref{fig:WenuraZoom1}(b), (c), and (d). Panel (b) shows a region where a small gap has not been closed, revealing Nb$_3$Sn grains that span the joint while being attached to either side. Panel (c) shows a more typical region with Nb$_3$Sn grains continuously covering the joint. In panel (d), the seam is mostly hidden by growth of Nb$_3$Sn grains across it.

\begin{figure}
    \centering
    \includegraphics[width=.8\textwidth]{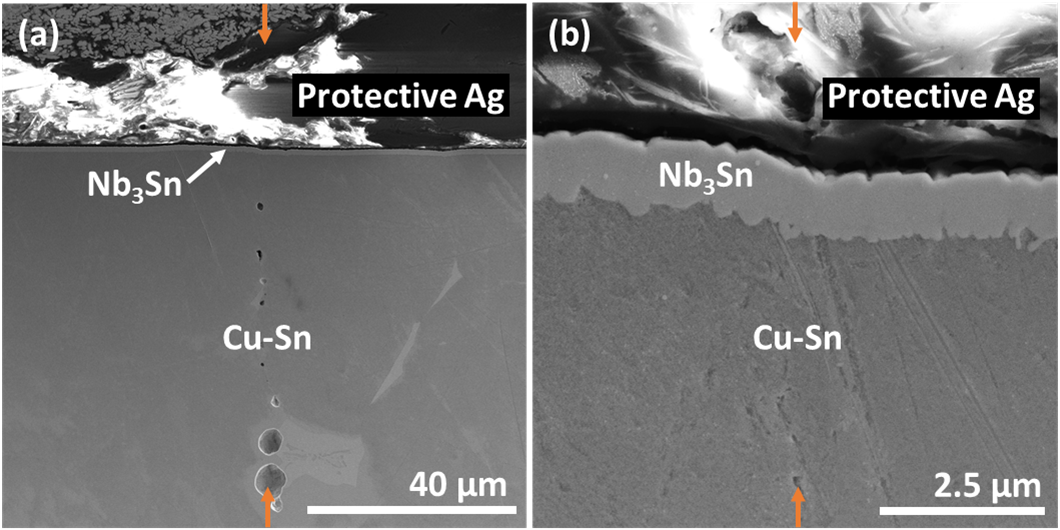}
    \caption[]{SEM cross-section images of the joint formed by the HB recipe. Image (a) shows a substantial region of the diffusion-bonded bronze joint, with the Nb$_3$Sn evident as a thin region near the image top. Large and small voids are seen along the joint interface. Image (b) magnifies the Nb$_3$Sn layer, showing continuity of the layer across the joint. Orange arrows in each panel guide to the seam.}
    \label{fig:CrossSectionContinuous}
\end{figure}

Images of the joint cross-section shown in Fig.~\ref{fig:CrossSectionContinuous}(a) and (b) verify the interdiffusion of bronze across the joint face and coverage of the joint with Nb$_3$Sn. In (a), much of the image shows the diffusion bond between the bronze blocks' joint faces. A few spherical voids have evidently coalesced, suggesting near completion of diffusion mechanisms active during bonding. These voids allowed geometric imperfections and surface defects to migrate from other areas of the joint to accumulate in the void. We infer that, since the temperature used for bonding the joint is approximately 80\% of the melting point of the bronze, high atomic mobility facilitated coalescence \cite{JMatSci2016}. The light gray regions between the voids span the joint, indicating re-growth of grains across the joint. In Fig.~\ref{fig:CrossSectionContinuous}(b), the Nb$_3$Sn film maintains even thickness across the joint while accommodating a small height mismatch at the region sampled for the image. As suggested in prior work \cite{Withanage2021}, Sn activity at the bronze surface likely was high enough for rapid conversion of the deposited Nb to Nb$_3$Sn. Since no change in the Nb$_3$Sn coating is evident above the joint, the hot Nb deposition and pre-heating bonding step may have prevented the joint from acting as a high-mobility pathway for Sn diffusion or Nb infiltration, mechanisms that obstructed solid-state diffusion between the bronze blocks in the other recipes.



\begin{figure}
    \centering
    \includegraphics[width=.5\textwidth]{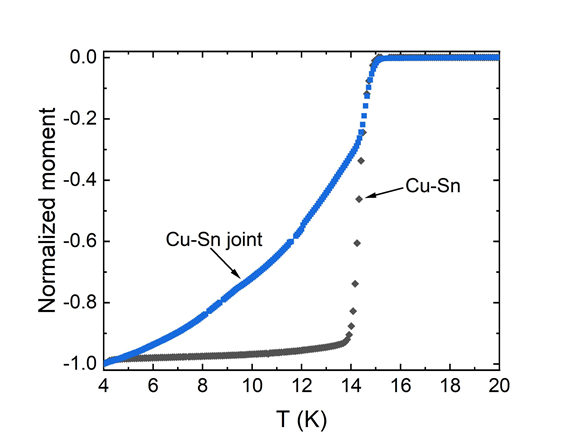}
    \caption{This plot shows magnetic moment vs temperature data for a Nb$_3$Sn film spanning a joint made by the HB recipe (labeled Cu-Sn joint)
    and for a film made on a solid bronze coupon (labeled Cu-Sn) with 2-hour longer reaction.}
    \label{fig:Tc}
\end{figure}

Figure~\ref{fig:Tc} shows T$_c$ measurements for the joint produced by the HB recipe in comparison to a film grown on a hot bronze substrate with no joint and a 3-hour (vs. 1 hour for the HB recipe) post-reaction duration. The superconducting transition onset is identical for Nb$_3$Sn films at $\sim$\qty{15}{\kelvin}, which is reduced from the ideal value of \qty{18.3}{\kelvin} \cite{Devantay1981} due to the strain resulting from CTE mismatch between Nb$_3$Sn and bronze \cite{Withanage2021, Easton1980}. The breadth of the transition for the specimen with the joint is probably due to incomplete conversion of Nb to stoichiometric Nb$_3$Sn, since the $T_c$ value decreases in proportion to the amount of Sn deficiency down to 18-at.\% Sn \cite{Devantay1981}. Because the HB recipe used a prior heating step to form the diffusion bond, the deposition chamber walls were warm at the time the Nb deposition began. Chamber temperature limits were reached when the reaction duration reached 1 hour, which prevented further reaction of this sample without an unusual cool-down and reheating step. A longer reaction after Nb deposition could be facilitated by different equipment to achieve the sharp transition shown for the Nb$_3$Sn film on a bronze piece without a joint. Nonetheless, the nearly duplicate temperature transition behavior near the transition onset suggests that the best material, which is the material next to the bronze, should be consistent.

\begin{figure}
    \centering
    \includegraphics[width=.8\textwidth]{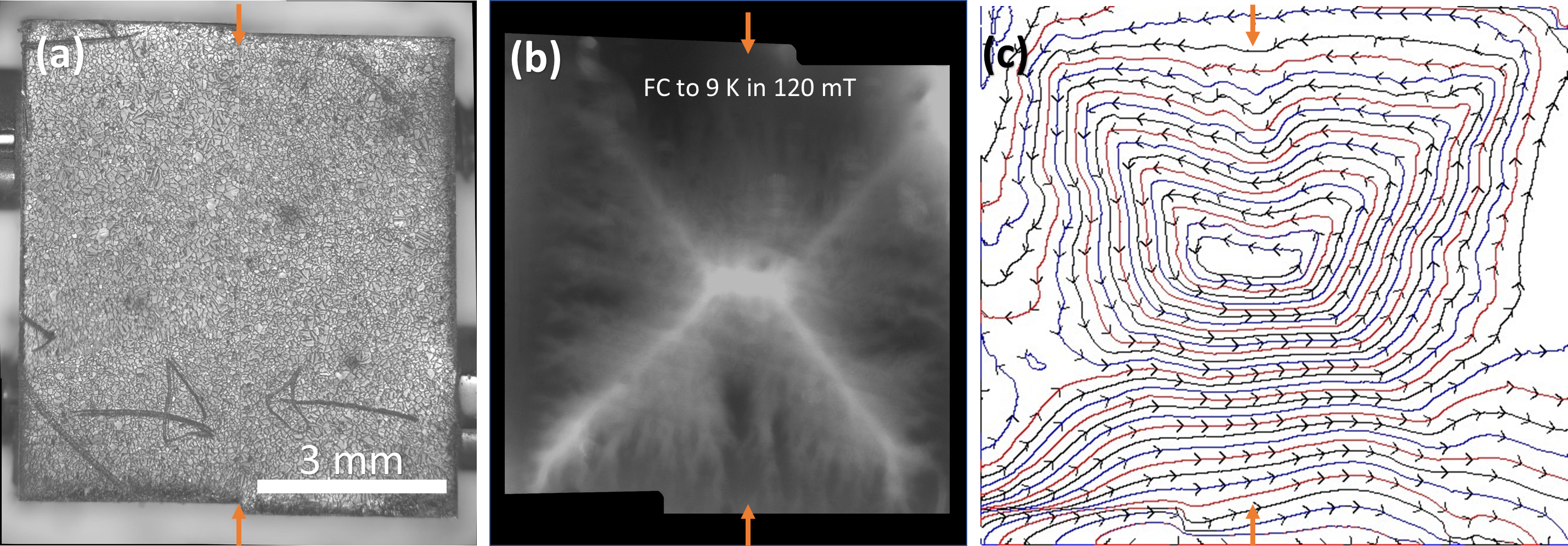}
    \caption{Panel (a) shows a plan-view image of the Nb$_3$Sn film grown on the common substrate face of joined bronze blocks using the HB recipe taken through the microscope of the MOI system. Orange arrows in this and the other panels indicate the position of the joint between the bronze blocks. Panel (b) shows the pattern created on the magnetic indicator film placed on top of the Nb$_3$Sn at a temperature of \qty{9}{\kelvin} after field cooling with a perpendicular field of \qty{140}{\milli\tesla}. Bright regions indicate higher magnetic flux, and the ``rooftop'' pattern seen is created by uniform screening currents in the Nb$_3$Sn film, as indicated in panel (c). }
    \label{fig:MOI}
\end{figure}

Further evaluation of the continuity of the HB joint used magneto-optical imaging (MOI), which is exquisitely sensitive to variations of supercurrent flow. MOI suggested no disruption of superconducting continuity, as shown in Fig.~\ref{fig:MOI}. MOI data was obtained at \qty{9}{\kelvin} in both the applied field and field-removed state where screening currents and trapped flux create strong contrast in the Faraday effect indicator film \cite{Polyanskii2021}. The sequence of panels in (a)-(c) do not suggest any disruption of screening current at the joint. MOI analysis of the most continuous post-reacted film (Recipe CB) showed no supercurrent across the seam, consistent with microscopy observations.

\subsection{Broader Implications}

It should be possible to adapt the recipes presented here to other techniques for forming Nb$_3$Sn on joined metal surfaces. An interesting potential advantage is the expansion of 3 moles of Nb (bcc lattice parameter \qty{0.330}{\nano\meter}) by 37\% to accommodate the addition of Sn atoms and the transformation to 1 mole of Nb$_3$Sn (A15 crystal structure with lattice parameter \qty{0.529}{\nano\meter}) via reactions using the bronze route \cite{Suenaga1986}. This was one factor that led us to investigate whether a Nb$_3$Sn film might span a joint between bronze pieces. Logically, the same advantage should apply for any reaction where a Nb coating is converted to Nb$_3$Sn via addition of Sn atoms from an external source. The use of bronze in the present work allowed temperatures of $\sim$\qty{715}{\degreeCelsius} to be used for the Nb-Sn diffusion reaction, and similar results might be expected for the Sn-vapor reaction methods at $\sim$\qty{1100}{\degreeCelsius}.

Cylindrical magnetrons are a convenient method to deposit metals on the inside of polished cavities. The results in the present work discourage application of Nb film deposition at low temperatures, where it was postulated that the CTE mismatch between Nb and bronze caused separation of the Nb and Nb$_3$Sn film across the joint region. The present results encourage the much more difficult approach using cylindrical magnetrons in hot oven-like surroundings, which presents several challenges including high cooling flow in a tiny cavity space to prevent demagnetization. Recent work by Valizadeh et al. at Daresbury~\cite{Valizadeh2025} demonstrated a viable hardware solution using planar magnetrons on the end of a stick, positioned on traveling stages inside the cavity. The divergence of the sputtered plume from planar magnetrons proved sufficient to coat cavity walls uniformly, while the planar geometry enabled efficient cooling of the magnetron. The Daresbury team successfully operated their system at \qty{580}{\degreeCelsius} for depositing Nb$_3$Sn from a stoichiometric target. Chemical vapor routes could also have advantages in hot deposition conditions provided that other challenges such as halogen reactions can be managed.

A reliable healing process for Nb$_3$Sn joints also hinges on careful substrate preparation with precise mechanical alignment. In the present work, while mirror-smooth polishing was achieved, flatness tolerances to prevent gaps between the joint faces were not considered. All of our bronze blocks exhibited growth of the bronze grains during heating, which caused small eruptions from the polished surface and revealed height contrast as exemplified in Fig.~\ref{fig:Recipe3Seam} and Fig.~\ref{fig:WenuraZoom1}.


\section{Conclusions}

In this study, we applied multiple processes to coat bronze pieces joined along a common seam with Nb and subsequently facilitate formation of a $\sim$\qty{1}{\micro\meter} thick film of Nb$_3$Sn that uniformly spans the seam. Best results were obtained for a ``hot bronze'' recipe \cite{Withanage2021}, where Nb deposition was carried out while the bronze was heated to $\sim$\qty{715}{\degreeCelsius}. Magneto-optical characterizations did not indicate adverse impacts on supercurrent flow across the seam, and Nb$_3$Sn grains had uniform size and morphology across the joint. Critical temperature was consistent with Nb$_3$Sn formed by the bronze route when taking into account the thermal contraction mismatch strain between bronze and Nb$_3$Sn. Less encouraging results were obtained when Nb was deposited at \qty{200}{\degreeCelsius}, and disconnections of the Nb$_3$Sn film from the bronze were observed in microstructural analyses. The results suggested that mismatch of the coefficient of thermal expansion between bronze and both Nb and Nb$_3$Sn may have been a cause of disconnections, even though adhesion of the Nb$_3$Sn to the bronze survived thermal cycles.

This work demonstrates the first continuous Nb$_3$Sn morphology across a joint for SRF cavity applications. Further development of the techniques used herein could serve as an alternative Nb$_3$Sn cavity fabrication route using half-shells or multiple pieces. This approach may also contribute toward development of persistent joints in magnet wires and connections for quantum devices. An important outcome of the MOI analysis was lack of evidence of blocking current at \qty{9}{\kelvin}, well above the operating temperature range for magnets, cavities, and quantum devices.

\ack{The authors thank Dr. Akhiro Kikuchi at NIMS, Japan, for providing Ti-doped bronze. We also thank Peter J. Lee (ASC-NHMFL), and Shreyas Balachandran (ASC-NHMFL) for their insights on this work.}

\funding{This work was supported by the U.S. Department of Energy, Office of Science, Office of High Energy Physics under Award No. DE-SC 0018379. A portion of this work was performed at the National High Magnetic Field Laboratory, supported by National Science Foundation Cooperative Agreement No. DMR-1644779 and the State of Florida.}

\roles{Andre Juliao: Investigation, Formal analysis, Visualization, Resources, Writing -- original draft, Writing -- review \& editing. Wenura Withanage: Conceptualization, Methodology, Investigation, Resources, Writing -- original draft, Writing -- review \& editing. Nikolya Cadavid: Investigation, Formal analysis, Writing -- review \& editing. Anatolii Polyanskii: Investigation, Formal analysis. Lance D. Cooley: Conceptualization, Funding acquisition, Resources, Supervision, Writing -- review \& editing.}

\data{The data that support the findings of this study are available upon reasonable request from the authors.}

\suppdata{}


\bibliography{bibfile.bib}

\end{document}